\documentclass[aps,prx, groupedaddress, twocolumn]{revtex4-2}

\usepackage{lineno}
\usepackage{xcolor}
\usepackage{graphicx}
\usepackage{dcolumn}
\usepackage{bm}
\usepackage[normalem]{ulem}

\newcommand {\be}{\begin{equation}}
\newcommand {\ee}{\end{equation}}
\newcommand {\bea}{\begin{eqnarray}}
\newcommand {\eea}{\end{eqnarray}}
\newcommand {\FIG}[1]{Fig. \ref{#1}}
\newcommand {\FIGS}[1]{Figs. \ref{#1}}
\newcommand {\EQ}[1]{Eq. (\ref{#1})}
\newcommand {\EQS}[2]{Eqs. (\ref{#1}) and (\ref{#2})}

\begin{document}


\title{A simple and efficient model for epidemic control on multiplex networks}


\author{Minsuk Kim}
\author{Soon-Hyung Yook}%
 \email{syook@khu.ac.kr}
\affiliation{%
 Department of Physics and Research Institute for Basic Sciences, Kyung Hee University, Seoul 130-701, Korea
}%


\date{\today}

\begin{abstract}

When an unprecedented infectious disease with high mortality and transmissibility emerges, immediate usage of vaccines or medicines is hardly available.
Thus, many health authorities rely on non-pharmaceutical interventions through traceable
fixed contacts.
However, in reality, there is an additional type of transmission routes to the regular
and fixed contacts: the random anonymous infection cases where non-pharmaceutical interventions are hardly feasible.
In our study, such realistic situations are implemented by 
the susceptible-infected-recovered model with isolation on
multiplex networks. The multiplex networks are composed of a fixed 
interaction layer and a layer with time-varying random interactions 
to represent the different types of disease spreading routes.
The multiplex networks represent the combinations of the quenched disorder and annealed disorder.
Here, we suggest a preemptive isolation protocol which isolates the second nearest neighbors 
of the hospitalized individuals and compare it with one of the most popular protocol
adopted by many health organizations over the globe.
From numerical simulations we find that our preemptive measure
significantly reduces both the final epidemic size and the number of the isolated per unit time.
Our finding suggests a better non-pharmaceutical intervention which can be adopted to various types of diseases
even though the contact tracing is only partially available.
\end{abstract}


\maketitle

\section{Introduction}
An outbreak of a new disease, such as the bubonic plague pandemic
in the 14th century \cite{gould_book}, 
the 1918 influenza pandemic \cite{taubenberger06},
and the recent outbreak of severe acute respiratory syndrome \cite{chan-yeung03},
has been a large threat throughout human history.
Despite great advances in medical science and pharmacology, 
immediate use of an effective vaccine or
antiviral drug is not always possible when {\it new} infectious diseases emerge.
For example, due to the absence of vaccines or antiviral drugs for new severe acute
respiratory coronavirus 2 (SARS-CoV-2)
during the early stage of the coronavirus disease 2019 (COVID-19) pandemic,
more than 172 million people have been infected
and has caused more than 3.7 million deaths until March 2021 \cite{coronaboard}.
Thus, finding an efficient non-pharmaceutical intervention (NPI) is crucial
to mitigate the pandemic situation for new emerging diseases.

The best NPI for a new disease is a perfect lockdown,
under which all individuals are strictly isolated.
For example, during the early stage of the COVID-19 pandemic, 
strict lockdown measures had been successfully applied in mainland China 
and many European countries \cite{xuefei20,guangyu21}.
However, the strict lockdown policy is not sustainable
if the pandemic period continues long enough to cause a severe recession of economic
activity and to increase social fatigue \cite{Nigel20,Maria20}.
Thus, it is necessary to find NPIs that can both suppress the epidemic spreading and minimize the negative impact on social and economic systems.
To meet these demands, various NPIs have been intensively studied based on real data and theoretical models to alleviate the recent pandemic situations \cite{Ferretti20,flaxman20,Perra21,Maier20,Chan21,thurner20,khain20,mukhamadiarov21,choi20,arenas20,schlosser20,Sneppen21,pastor-satorras01,newman02,Balcan09,Zuzek15}.

Among the various NPIs, the quarantine of the infected individuals and their contacts
is one of the most intuitive measures and commonly shared by many health authorities over
the world \cite{wendy20}.
Thus, the isolation of the infected \cite{Zuzek15,arenas20} and
tracing the contacts \cite{Fraser04,Peak17,Ferretti20} are two important factors for epidemic control problems.
However, if the infections from asymptomatic and pre-symptomatic patients are
potential transmission routes like the COVID-19 case \cite{Tong20,Bai20},
finding the contacts with such patients is not trivial.
Furthermore, when airborne transmission is another
important route of spreading \cite{Zhang20},
the tracing becomes much harder due to the random anonymous contacts
through the publicly opened environments \cite{Sneppen21}.
In this study, such random anonymous transmission is implemented by
the double-layered multiplex networks (DLMNs) \cite{newman_book}.
To consider the situation with the pre-symptomatic and asymptomatic transmissions, we assume that 
individuals in our models have one of the following disease states, susceptible ($S$),
infected ($I$), and recovered ($R$) \cite{herbert00}.
Furthermore, the contact tracing probability and isolation states are introduced
to account for more realistic situations in our model.
As we will show, we first model the most popular protocol for the NPI
adopted by many health authorities, and also introduce a reinforced
protocol model. From the quantitative comparison of
the two models, we suggest a simple and efficient
NPI strategy for epidemic control of any emerging infectious disease
by only using the known topology of the fixed interaction layer.

\section{Model}

\subsection{Construction of the double-layered multiplex networks}

\begin{figure}[ht]
    \centering
    \includegraphics[width=0.45\textwidth]{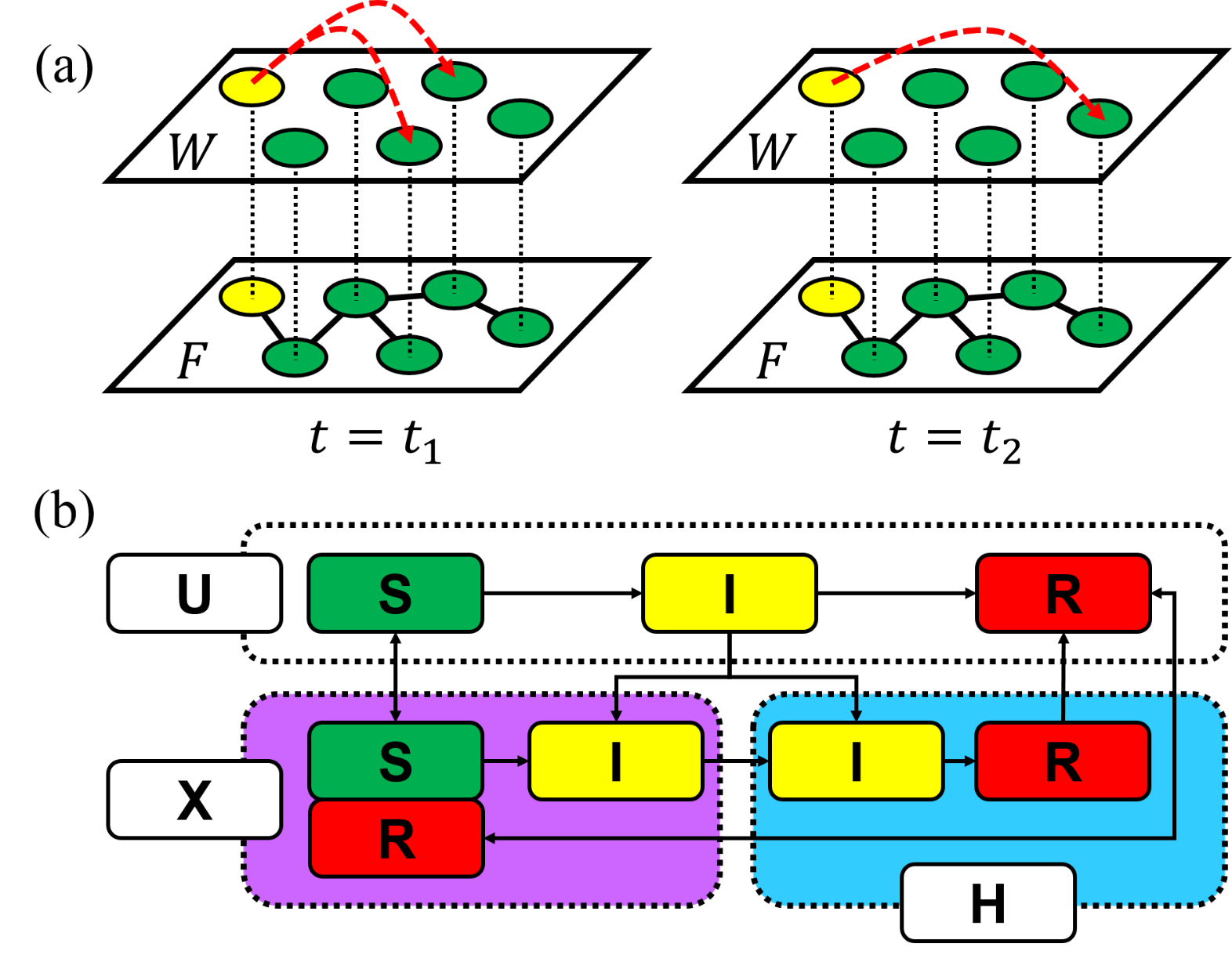}
    \caption{(a) The schematic diagram for the interaction on the DLMN.
    $W$ and $F$ layers are composed of the same set of nodes, i.e.,
    the nodes at the ends of each dotted line are identical.
    The yellow infected node chooses random partners on $W$ at each time step $t=t_1$ and $t=t_2$ (red dashed arrows).
    Thus, on $W$ it interacts with different nodes at each $t$,
    while its interacting partner does not change on $F$ (black solid lines).
    (b) The change of states under the BIP or RIP.
    Red, yellow, and green boxes denote the states for $\sigma_1$, and white boxes
    represent the states for $\sigma_2$. }
    \label{scheme}
\end{figure}

Individuals are denoted by nodes and the interactions between them are represented
by links in the DLMN.
Let $F$ and $W$ be the two layers in the DLMN (see \FIG{scheme} (a)). 
On $F$ each node is connected with randomly chosen $k$ nodes drawn 
from a given degree distribution $P_{F}(k)$. 
The topology of the network on $F$ does not change in time, which
corresponds to the quenched structural disorder.
At the same time, each node interacts with $k^\prime(t)$ random nodes on $W$,
where $k^\prime(t)$ is drawn from another degree distribution $P_{W}(k^ \prime)$ at each time $t$.
Thus, the interaction topology on $W$ changes with $t$, which represents
the annealed structural disorder.
Under severe epidemic situations, the government tries to cordon people
off public facilities, and each individual refrains from social activities.
Thus, the number of contacts of each individual
is significantly restricted and homogeneous.
To generate such homogeneous interaction structures,
we use the Poisson distribution, 
$P(k)=\left<k\right>^{k}e^{-\left<k\right>}/k!$,
for both $P_F(k)$ and $P_W(k^\prime)$ 
\cite{newman_book,Solomonoff51,Erdos60}.
Here $\left<k\right>$ is the mean degree of the network.
This model can be easily extended to any interaction topology.
For example, the results for $P_F(k)\sim k^{-\gamma}$
on the $F$-layer are also displayed in the Appendix. Here $\gamma$ is
the degree exponent.

The construction of each layer is as follows. 
Let $N$ be the number of nodes in a network.
To construct a fixed random network on $F$, we randomly select two nodes among $N$ nodes
and connect them if they are not linked. This process continues until we
have $L=N\left<k\right>/2$ links on the $F$ layer. 
The degree distribution of the 
obtained network on $F$, $P_F(k)$, is known to be the Poisson distribution. 
On the other hand, the topology of the network on $W$ changes with time.
Therefore, at each time $t$, an infected node $i$ randomly
chooses $k_i^\prime$ neighbors on $W$. $k_i^\prime$ is drawn from the
Poisson distribution at each $t$.

\subsection{Intervention strategies}

The state of each node at $t$ in the DLMN is 
described by a two-component variable 
${\bm \sigma}=(\sigma_{1},\sigma_{2})$.
$\sigma_{1}$ has one of the three disease states: $S$,
$I$, and $R$. 
$\sigma_{2}$ denotes the state for isolation measure.
Here, three isolation states are possible: 
i) {\it self-isolation} at home when an individual feels mild symptoms or
recognizes a suspicious contact but has not been confirmed yet,
and ii) {\it hospitalization} by the health authority
when the patient is confirmed to be infected.
If an individual is not isolated then it is in the iii) {\it unisolated} state.
Thus, $\sigma_2$ can be one of the following states: self-isolated ($X$),
hospitalized ($H$), and unisolated ($U$) (see \FIG{scheme} (b)).

Since we cannot trace the contacts on $W$ due to the random anonymity,
the self-isolation for the suspicious contacts
can be applied only to $F$.
Depending on the range of the self-isolation, we introduce two intervention
protocols, the basic isolation protocol (BIP) and the reinforced isolation protocol (RIP).
Under the BIP only the confirmed patients and the one who has direct contact with
the confirmed patient on $F$ are isolated. This is the most popular
quarantine protocol adopted by many health authorities \cite{webster20}.
However, the fraction of the household or workplace infection cannot be ignored in some infectious 
diseases, for example, the household infection is more than 15\% for COVID-19 \cite{Park20}.
Such household and workplace infection can be caused by a self-isolated individual.
Furthermore, such transmissible paths can become a part of super-spreading events
if the pre-symptomatic or asymptomatic infection is possible.
Thus controlling such local contacts is another important factor to mitigate the transmission.
For a preemptive protection of the susceptible, in the RIP model if a node is hospitalized,
then its first and also second nearest neighbors on $F$ are isolated.

\subsection{Basic isolation protocol}
\begin{figure}[h]
    \centering
    \includegraphics[width=0.4\textwidth]{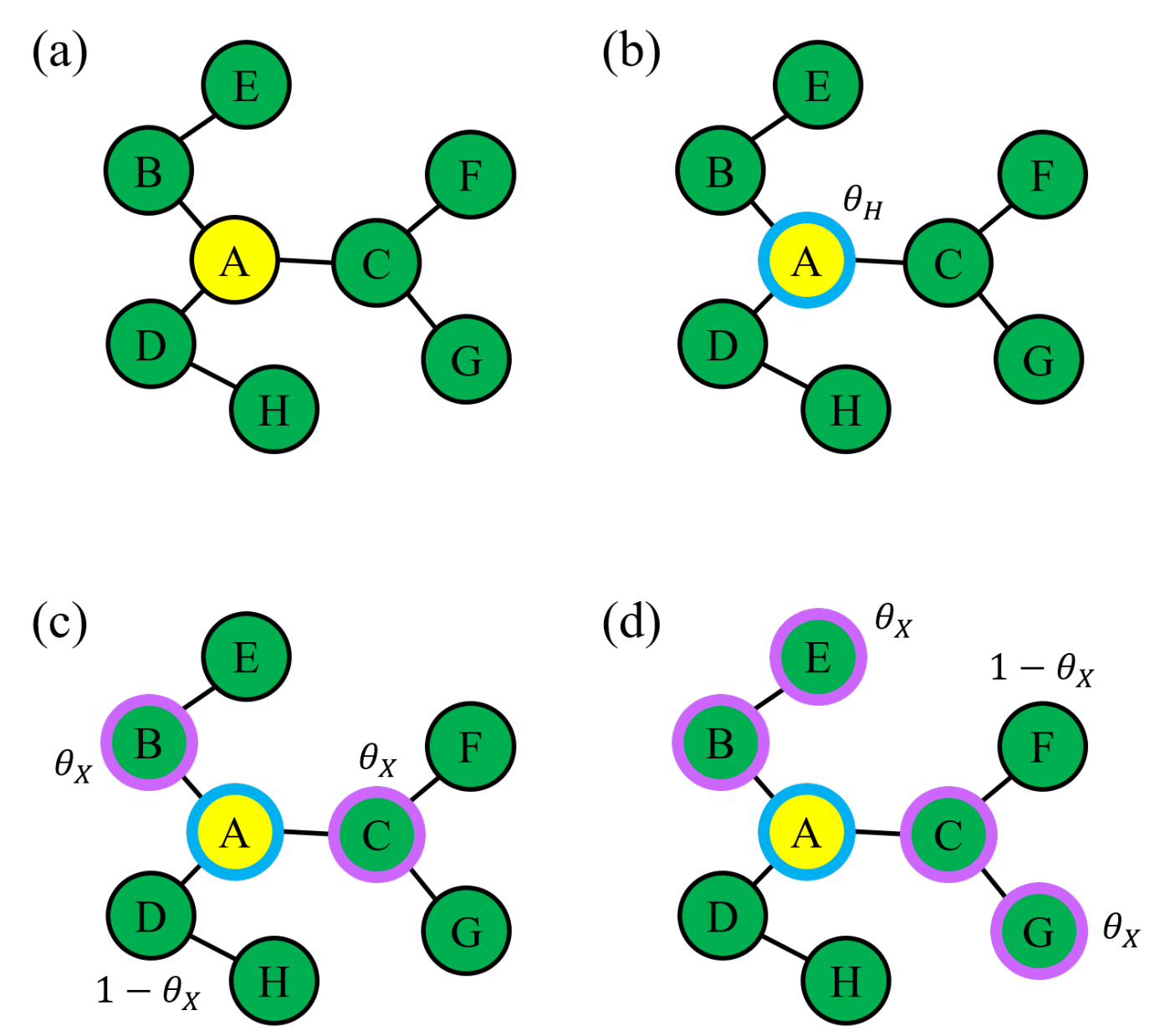}
    \caption{(a) On a network with size $N=8$ the node $A$ is 
infected while nodes $B\sim H$ are susceptible.
(b) Node $A$ is hospitalized with probability $\theta_{H}$.
(c) BIP and RIP: Node $B$ and $C$ are self-isolated with 
probability $\theta_X$ while node $D$ remains unisolated with
probability $1-\theta_X$.
(d) RIP: In addition to (c) node $E$ and $G$ are self-isolated
with probability $\theta_X$ while node $F$ remains unisolated with
probability $1-\theta_X$. Node $H$ are not included in the
self-isolation candidate since node $D$ is not self-isolated
in step (c).}
    \label{protocol}
\end{figure}
In our models, each infected node transmits the disease to the connected susceptible nodes with
the probability $\beta_F$ ($\beta_W$) on $F$ ($W$).
With the probability $\theta_X$ ($\theta_H$) the nodes are
self-isolated (hospitalized) for the isolation (hospitalization) period $t_X$ ($t_H$).
The infected nodes are recovered after the recovery time $t_R$.
To specify the update rule for each protocol, we introduce additional parameters
$T_i^I$, $T_i^X$, and $T_i^H$ which denote the time of infection, self-isolation,
and hospitalization for each node $i$, respectively.

In the BIP model, all nodes are initially in the state ${\bm \sigma}=(S,U)$.
Then a node $i$ is randomly selected and set to be ${\bm \sigma}_i=(I,U)$ and $T^I_i=0$.
At each time step $t$, three processes are repeated for all infected nodes whose time
of infection is $T^I<t$:
(1) {\it infection}, (2) {\it isolation},
and (3) {\it unisolation and recovery}.
Each process is composed of the following sub-processes.
{\it Infection}: (1-i) Each node $i$ with ${\bm \sigma}_i=(I,U)$ or $(I,X)$
transmits the disease to $k_{i}$ connected
nodes on $F$ with the probability $\beta_F$ if the state of the connected node $j$
is ${\bm \sigma}_{j}=(S,U)$ or $(S,X)$.
(1-ii) If $\sigma_{i,2}=U$, then it randomly chooses $k^\prime_{i}$ nodes on $W$  
and infects with probability $\beta_W$ when the randomly chosen node $j$ is in the state ${\bm \sigma}_j=(S,U)$.
$T^I_j$ for all new infected nodes $j$ is set to be $T^I_j=t$.
{\it Isolation}: (2-i) Each infected node $i$ with $T^I_i<t$ is hospitalized with
the probability $\theta_{H}$, i.e., $\sigma_{i,2}=H$ and $T^H_i=t$.
(2-ii) Let $\Gamma_i$ be the set of nodes connected to the hospitalized node $i$ on $F$.
Then we set $\sigma_{j,2}=X$ and $T^X_j=t$ for all $j \in \Gamma_i$
with the probability $\theta_{X}$, if $\sigma_{j,2}=U$ at $t-1$. 
This corresponds to the self-isolation.
(2-iii) If the state of node $i$ was ${\bm \sigma}_i=(I,X)$ at $t-1$,
then it becomes ${\bm \sigma}_i=(I,H)$ and $T^H_i=t$ for all $i(=1,2,\cdots,N)$.
{\it Unisolation and recovery}: 
(3-i) For all nodes $i$ with $\sigma_{i,1}=I$ becomes $\sigma_{i,1}=R$,
if $t>T^I_i+t_{R}$. Here $t_R$ is a constant representing a recovery time from the infection.
(3-ii) For all nodes $i$ with $\sigma_{i,1}\in\left\{S,R\right\}$ and $\sigma_{i,2}=X$ 
are unisolated if $t>T_i^X+t_{X}$, where $t_X$ is the duration time for self-isolation.
(3-iii) For all nodes $i$ with ${\bm \sigma}_i=(R,H)$, 
if $t>T^H_i+t_H$ then the node $i$ is released from hospitalization,
and its state becomes ${\bm \sigma}_i=(R,U)$.
Here $t_H$ denote the duration time for hospitalization.
These processes are repeated until there left no infected node.
Under the BIP only the confirmed patients and the one who has direct contact with
the confirmed patient are isolated as shown in
\FIG{protocol}.
This is the most popular
quarantine protocol adopted by many health authorities over the globe \cite{webster20}.\\

\subsection{Reinforced isolation protocol}
The RIP can be implemented 
by adding the sub-process (2-iv) to the end of the
{\it isolation} process of the BIP: (2-iv) Let $j$ be the node
whose state is changed into $\sigma_{j,2}=X$ at $t$ and $\Gamma_j$ be the set of nodes
connected to $j$ on $F$. Then the state of node $n (\in \Gamma_j)$ with $\sigma_{n,2}=U$
also becomes $\sigma_{n,2}=X$ with the same probability $\theta_X$ for all $n$.
Thus, in the RIP model if a node is hospitalized,
then its first and second nearest neighbors on $F$ are isolated with the given probability (see \FIG{protocol}).

In the following simulations, we set the size of each layer as $N=100,000$
and use the mean degrees $\left< k\right>=\left<k^\prime\right>=8$.
The value of the mean degree only affects the epidemic
threshold \cite{newman_book}, and does 
not change the main conclusion.
The epidemic threshold of susceptible-infected-recovered (SIR) model is
related to percolation threshold and the branching factor \cite{newman_book, newman02, Zuzek15, Lagorio11}.
Since we are interested in the control of the severe epidemic outbreak,
we use $\beta_F=\beta_W\equiv \beta(=0.2)$ and $t_R=6$ to guarantee that the whole 
system becomes infected without any intervention (see Appendix \ref{epidemicthreshold}).
In our model, the strength of the intervention measures is controlled by
four parameters, $\theta_{H}$, $t_{H}$, $\theta_{X}$, and $t_{X}$.
For simplicity, we assume that $\theta_{H}=\theta_{X}\equiv\theta^*$
and $t_{H}=t_{X}\equiv t^*$.
Thus, we use only two control parameters $\theta^*$ and $t^*$.

\section{Results}
\subsection{The fraction of nodes in each state}
\begin{figure*}[ht]
    \centering
    \includegraphics[width=0.9\textwidth]{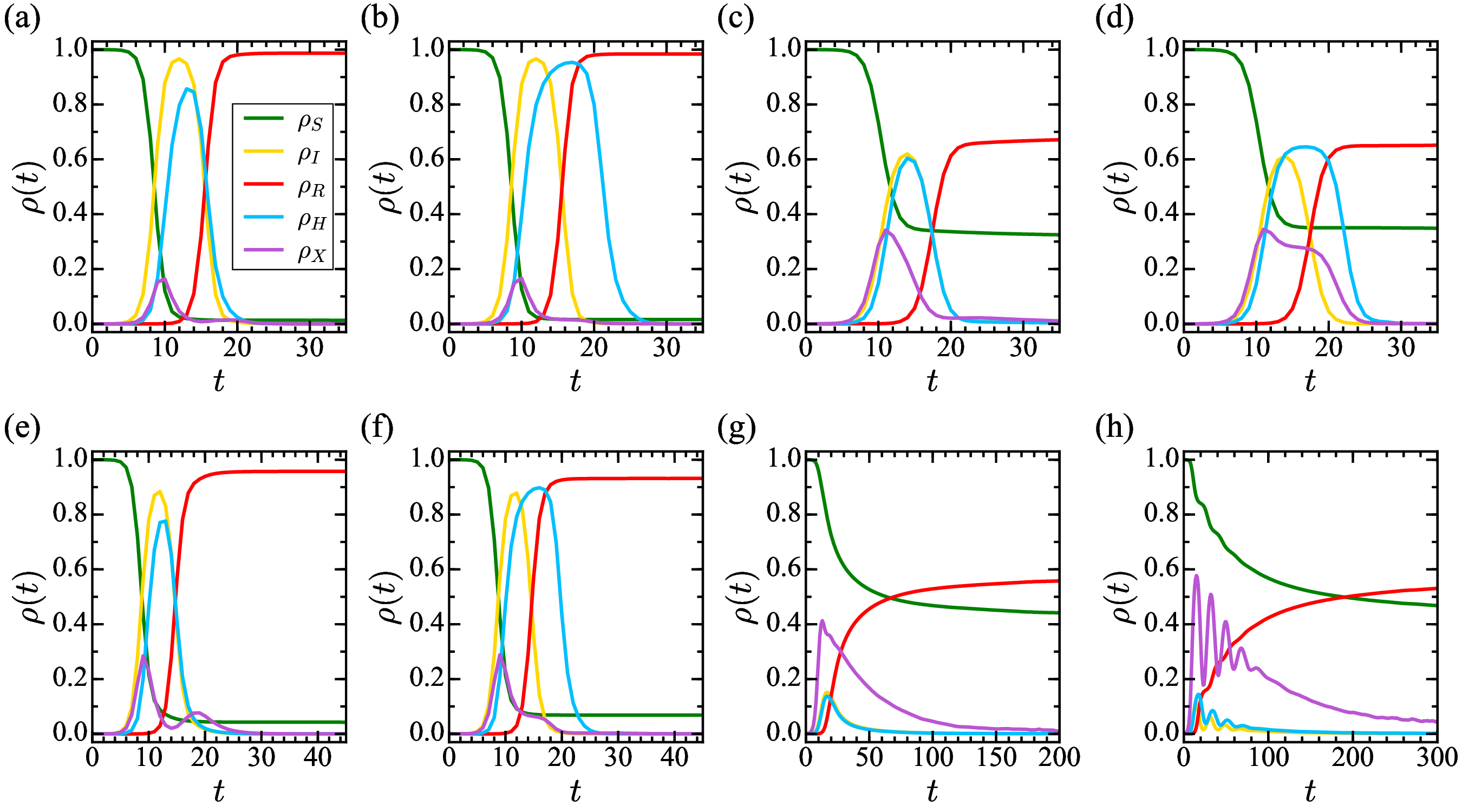}
    \caption{(a)-(d) are \{$\rho_m(t)$\}'s ($m\in \left\{S,I,R,H,X\right\}$) under the BIP
    with (a) $\theta^*=0.3$, 
    $t^*=4$, (b) $\theta^*=0.3$, $t^*=10$, (c) $\theta^*=0.9$, $t^*=4$,
    (d) $\theta^*=0.9$, $t^*=10$. 
    (e)-(h) show \{$\rho_m(t)$\}'s under the RIP with 
    (e) $\theta^*=0.3$, $t^*=4$, (f) $\theta^*=0.3$, $t^*=10$,
    (g) $\theta^*=0.9$, $t^*=4$, (h) $\theta^*=0.9$, $t^*=10$.
    $\beta=0.2$ and $t_{R}=6$ are used in common.
    Each data is obtained from 500 independent simulations by
    averaging the surviving samples at time $t$.}
    \label{rhos}
\end{figure*}
Let $\rho_m(t)$ ($m\in\{S,I,R\}$) be the fraction of nodes whose disease states is $\sigma_1=m$ at
time $t$, regardless of $\sigma_2$.
If $m\in\{U,H,X\}$, then $\rho_m(t)$ represents the fraction of nodes with $\sigma_2=m$.
The peak of $\rho_m(t)$ for each state $m$ is denoted by $\rho^{peak}_m$. 
In \FIGS{rhos} (a)-(d) we show $\{\rho_m(t)\}$'s under the BIP for various values of parameters $(\theta^*, t^*)$.
For small $\theta^*(=0.3)$, we find that $\rho^{peak}_I(t)>0.9$ followed by
$\rho^{peak}_H (\gtrsim 0.8)$, regardless of $t^*$ (\FIGS{rhos} (a), (b)).
$\rho_S(t)$ rapidly decreases and reaches $\rho_S\approx 0$ for $t>11$.
The value of $\rho^{peak}_X$ ($<0.2$) is relatively small.
Thus, $\rho_R(t\rightarrow\infty)\simeq 1$ when $\theta^*$ is small.
On the other hand, as $\theta^*$ increases, $\rho^{peak}_{I}$ is drastically suppressed 
as well as $\rho^{peak}_H$ (\FIGS{rhos} (c), (d)).
As a result, $\rho_R(t\rightarrow\infty)$ is reduced to $\rho_R(t\rightarrow\infty) \simeq 0.6\sim 0.7$ for $\theta^*=0.9$.
For both values of $\theta^*$ displayed in \FIGS{rhos},
$t^*$ only affects the behavior of $\rho_H$ and $\rho_X$ (the population of the isolated nodes).
Note that $\rho^{peak}_H$ is comparable with $\rho^{peak}_I$ for all values of
$(\theta^*,t^*)$, and $\rho_H(t)$ becomes wider as $t^*$ increases.
This means that the hospitalized period becomes longer without any significant
change in the final epidemic size, $\rho_R(t\rightarrow\infty)$, as $t^*$ increases for all $\theta^*$.
Thus, increasing $t^*$ without the improvement of traceability
causes an overload on the medical system by making patients
be hospitalized for a longer period.

\FIGS{rhos} (e)-(h) show $\{\rho_m(t)\}$'s for the RIP model.
When $\theta^*\lesssim0.3$, $\{\rho_m(t)\}$'s for the RIP model show
almost the similar behavior with those for the BIP, but $\rho_S(t\rightarrow\infty)$ for the RIP
is slightly larger than that for the BIP.
Since additional nodes
are self-isolated under the RIP, $\rho_{X}^{peak}$
increases compared with that for the BIP with the same $(\theta^*$, $t^*$).
However, we find that $\rho_H$ for the RIP becomes much smaller than that for the BIP.
This effect becomes more drastic for $\theta^*>0.3$.
For example, $\rho_R(t\rightarrow\infty)$ and $\rho_I^{peak}(t)$
significantly decrease to $\rho_R(t\rightarrow\infty)=0.5\sim 0.6$ and $\rho_I^{peak}\approx 0.1$
for the RIP with $\theta^*=0.9$.
The results indicate that the collapse of the medical systems
can be avoided under the RIP if we trace the contacts with sufficiently high accuracy.
In addition, we find $\rho_I$, $\rho_H$, and $\rho_X$
oscillate with decreasing amplitude under the RIP as $\theta^*$ increases.
This suggests that even though there is a rapid decrease in $\rho_I(t)$ after
its first peak when $\theta^*$ is sufficiently large,
it is still possible to be followed by successive multiple peaks of $\rho_I$.
See Appendix \ref{Oscillatory} for a more detailed description on this oscillatory behavior.
For comparison, we also display the evolution of $\left\{\rho_m(t)\right\}$'s
when $P_F(k)\sim k^{-\gamma}$ in Appendix \ref{rho_sfn}.
\subsection{The effective reproduction number}
\begin{figure*}[ht]
    \centering
    \includegraphics[width=0.85\textwidth]{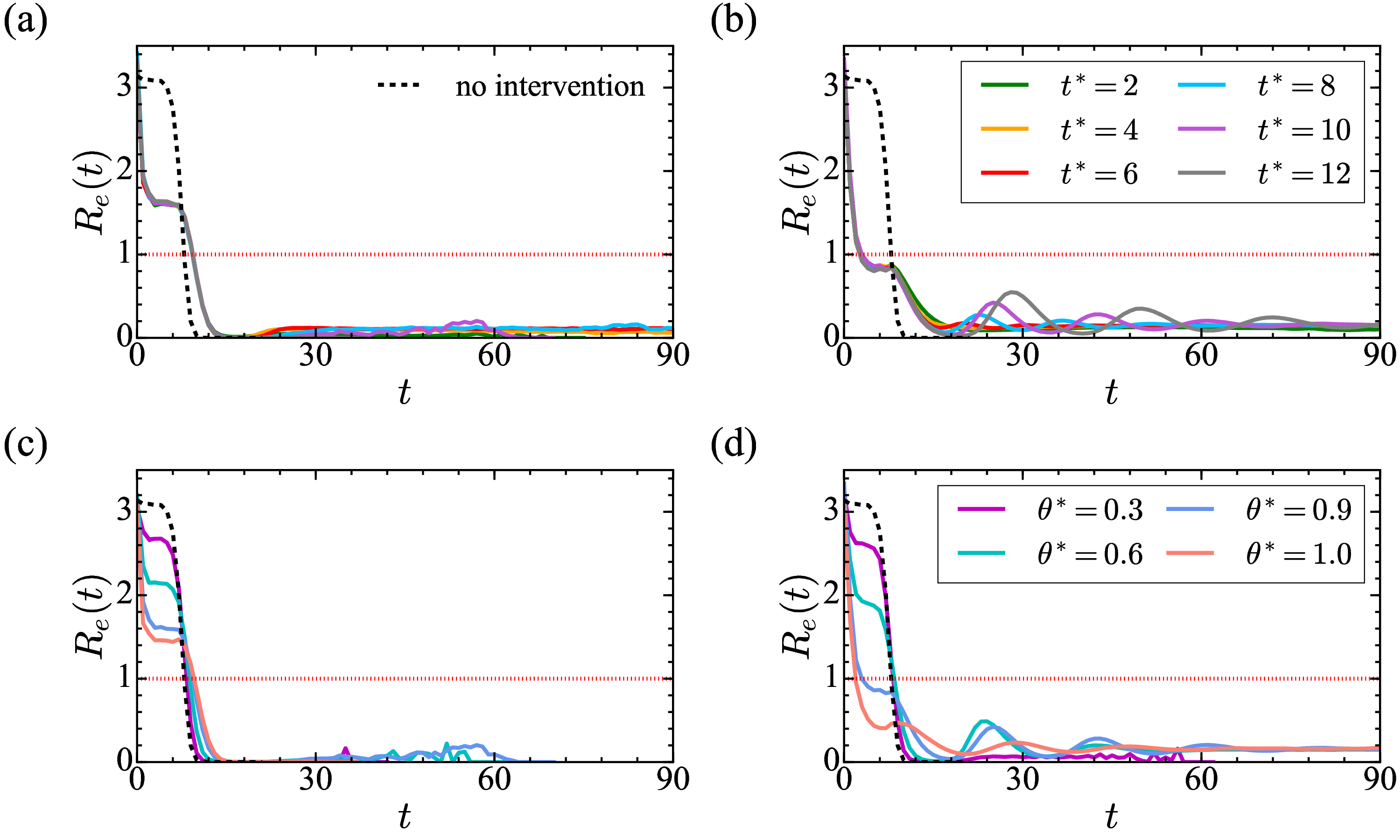} 
    \caption{Plot of $R_{e}(t)$ under the (a) BIP and (b) RIP when $t^*=2\sim 12$
and $\theta^*=0.9$ with $\beta=0.2$, $t_{R}=6$.
Plot of $R_{e}(t)$ for the (c) BIP and (d) RIP when $t^*=10$ and
$\theta^*=0,3\sim 1.0$ with $\beta=0.2$, $t_{R}=6$. 
The black dashed curve denotes the case when the isolation protocol is absent.
The red dotted horizontal line depicts $R_{e}(t)=1$.}
    \label{Rt}
\end{figure*}
To quantify the efficacy of intervention measures,
we estimate the instantaneous effective reproduction
number, $R_e(t)$, at $t$. For a practical purpose, we define $R_e(t)$ as
\begin{equation}
    R_e(t)=\frac{N^{new}_{I}(t)}{N_{I}(t-1)},
    \label{eq1}
\end{equation}
where $N^{new}_{I}(t)$ is the number of new infected nodes at $t$ and $N_I(t)$ is
the number of infected nodes at $t$ \cite{christophe07,anne13}. 
Thus $R_e(t)$ represents a metric to
quantify how many nodes are newly infected by the existing infected nodes at each $t$.

In \FIGS{Rt} (a) and (b), we show $R_e(t)$ for the BIP and RIP
with $\theta^*=0.9$ and $t^*=2\sim 12$.
The dashed line denotes $R_e(t)$ without intervention. 
As shown in  \FIG{Rt} (a), 
$R_e(t)$ for the BIP rapidly decreases when $t\lesssim 4$ 
and shows a plateau followed by another rapid drop, 
regardless of $t^*$. $R_e(t)=1$ at $t\approx 11$ 
and $R_e(t)<1$ for $t>11$.
When $t>15$ $R_e(t)$ approaches to $R_e(t)\approx 0$.
On the other hand, $R_e(t)$ for $\theta^*=0.9$ under the RIP
decreases more drastically and $R_e(t)<1$ for $t\gtrsim 5$ 
as shown in  \FIG{Rt} (b). 
When $t>20$, $R_e(t)$ oscillates with decreasing
amplitudes and approaches $R_e\approx 0$ under the RIP.

To investigate how $\theta^*$ affects 
the epidemic spreading, we also measure $R_e(t)$'s
for various $\theta^*$ when $t^*$ is fixed.
In \FIGS{Rt} (c) and (d), as an example, we display $R_e(t)$'s for $t^*=10$.
Since $\theta^*$ denotes traceability, 
$R_e(t)$ should decrease as $\theta^*$ increases for both protocols
as shown in \FIGS{Rt} (c) and (d). 
Note that when $\theta^*<0.3$, the difference between the BIP and RIP
is not noticeable.
However, if $\theta^*>0.3$, then $R_e(t)$ 
for the RIP becomes much smaller than those for the BIP.
From the data in \FIG{Rt}, we find that increasing $\theta^*$
is more important than increasing $t^*$.

The rapid drop of $R_e(t)$ under both protocols has two different
origins depending on $\theta^*$. For $\theta^*<0.3$ 
due to the large infection of the early stage, 
there does not remain a sufficient number of susceptible nodes
for $t>11$ (see \FIGS{rhos} (a), (b), (e), (f)). 
On the other hand, if $\theta^*>0.3$ then 
a significant amount of the susceptible is 
self-isolated, which protects the susceptible nodes before contact with 
the patients for $t>10$ (see \FIGS{rhos} (c), (d), (g), (h)). 
\subsection{The final epidemic size}
\begin{figure}[h]
    \centering
    \includegraphics[width=0.4\textwidth]{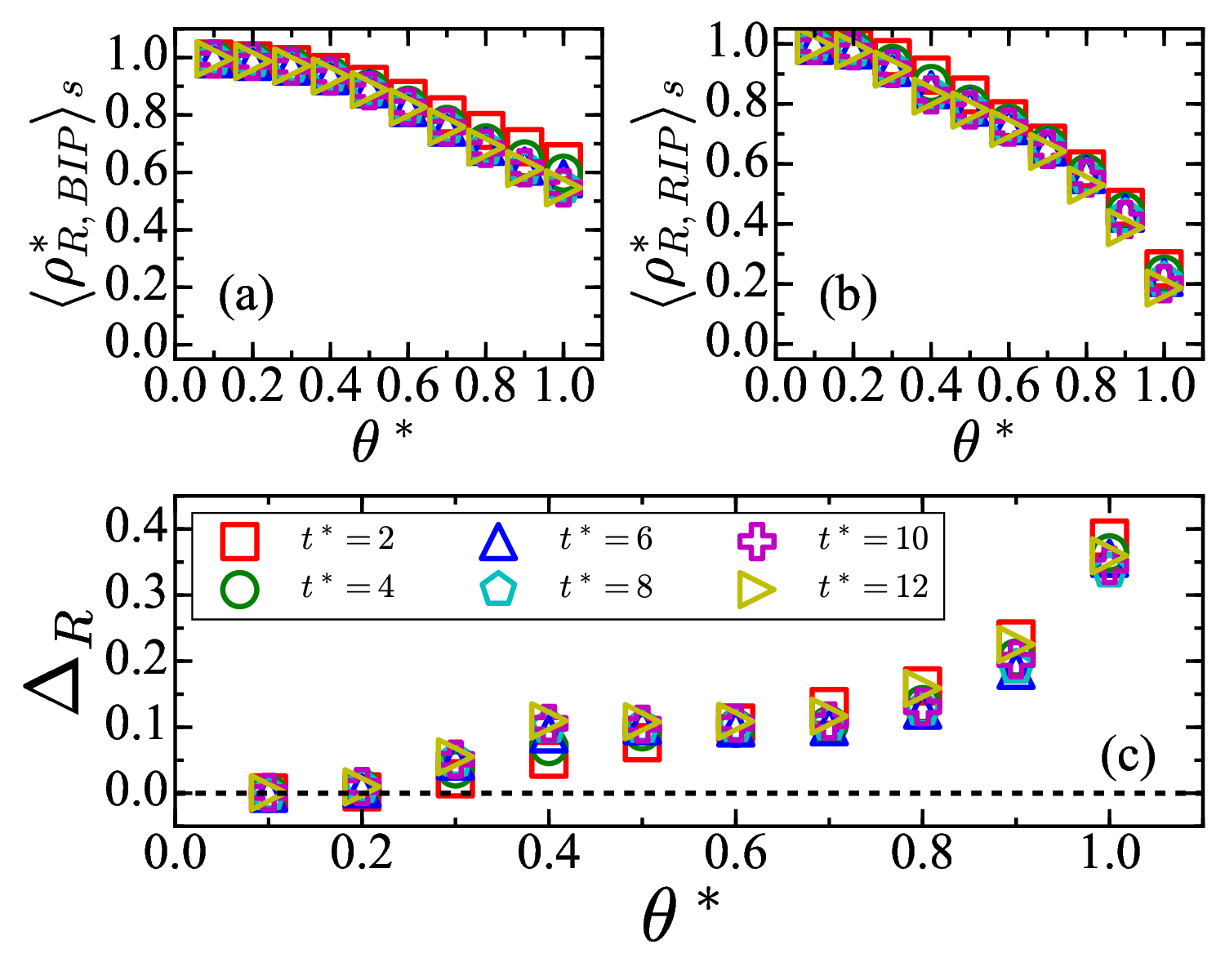}
    \caption{We plot the average final epidemic size under the
    (a) BIP and (b) RIP for $t^*=2\sim 12$ and $\theta^*=0.1\sim 1.0$.
    The increment of the isolation period ($t^*$) has negligible 
    effect on the final epidemic size while the traceability ($\theta^*$) significantly changes the final epidemic size.
    (c) We plot the difference of the final epidemic sizes between the BIP and RIP, $\Delta_{R}$. The black horizontal dashed line denotes
    $\Delta_{R}=0$.}
    \label{FinalSize}
\end{figure}
In order to evaluate the effectiveness of the isolation protocols, we 
obtained the final epidemic size. The final epidemic size under each
protocol $Y$($=BIP$ or $RIP$) is defined as
$\rho^*_{R,Y}\equiv\rho_{R,Y}(t\rightarrow\infty)$.
In \FIGS{FinalSize} (a) and (b), we plot $\left<\rho_{R,BIP}^{*}\right>_{s}$ and
$\left<\rho_{R,RIP}^{*}\right>_{s}$ with various parameter sets, $(\theta^*, t^*)$.
Here $\left<...\right>_{s}$ denotes the sample average over independent runs.
We used $10,000$ samples to obtain the average final epidemic size.
The results show that $t^{*}$ has a negligible effect on the final 
epidemic size. This suggests that extending the isolation period will simply add a burden on the socio-economic system unless there is any improvement in the ability to trace the infection routes. 

For a direct comparison between the two intervention protocols,
we measure the difference of the final
epidemic sizes between the BIP and RIP, $\Delta_{R}(t^*,\theta^*)$,
for each $t^*$ and $\theta^*$.
$\Delta_{R}(t^*,\theta^*)$ is defined as 
\bea
\Delta_{R}(t^*,\theta^*)\equiv \left<\rho_{R,BIP}^{*}(t^*,\theta^*)\right>_{s}-\left<\rho_{R,RIP}^{*}(t^*,\theta^*)\right>_{s}.
\eea
Thus, if $\Delta_{R}>0$ then $\rho_R$ for the BIP is larger
than that for the RIP.
The data in \FIG{FinalSize} (c) clearly shows that $\Delta_R$ rarely depends on $t^*$.
However, $\Delta_R$ strongly depends on $\theta^*$.
$\Delta_R\le 0.05$ for all $t^*$ when $\theta^*<0.3$,
while $\Delta_R>0.1$ for $\theta^*\gtrsim 0.4$ and $\Delta_R$ increases as $\theta^*$ increases.
This means that the RIP significantly reduces the final epidemic size 
compared to the BIP when $\theta^*\ge 0.3$.
For the maximal traceability, $\theta^*=1$,
we find that $\rho_R(t\rightarrow\infty)$ for the RIP is reduced
by 67\% compared to that for the BIP.
This corresponds to the 100\% increase of $\rho_S(t\rightarrow\infty)$ 
under the RIP compared to the BIP.
Thus, isolation of the possible suspicious contacts in advance by 
applying the RIP significantly reduces 
the final epidemic size when $\theta^*\ge0.3$.

\subsection{The number of isolated nodes per unit time}
\begin{figure}[h]
    \centering
    \includegraphics[width=0.4\textwidth]{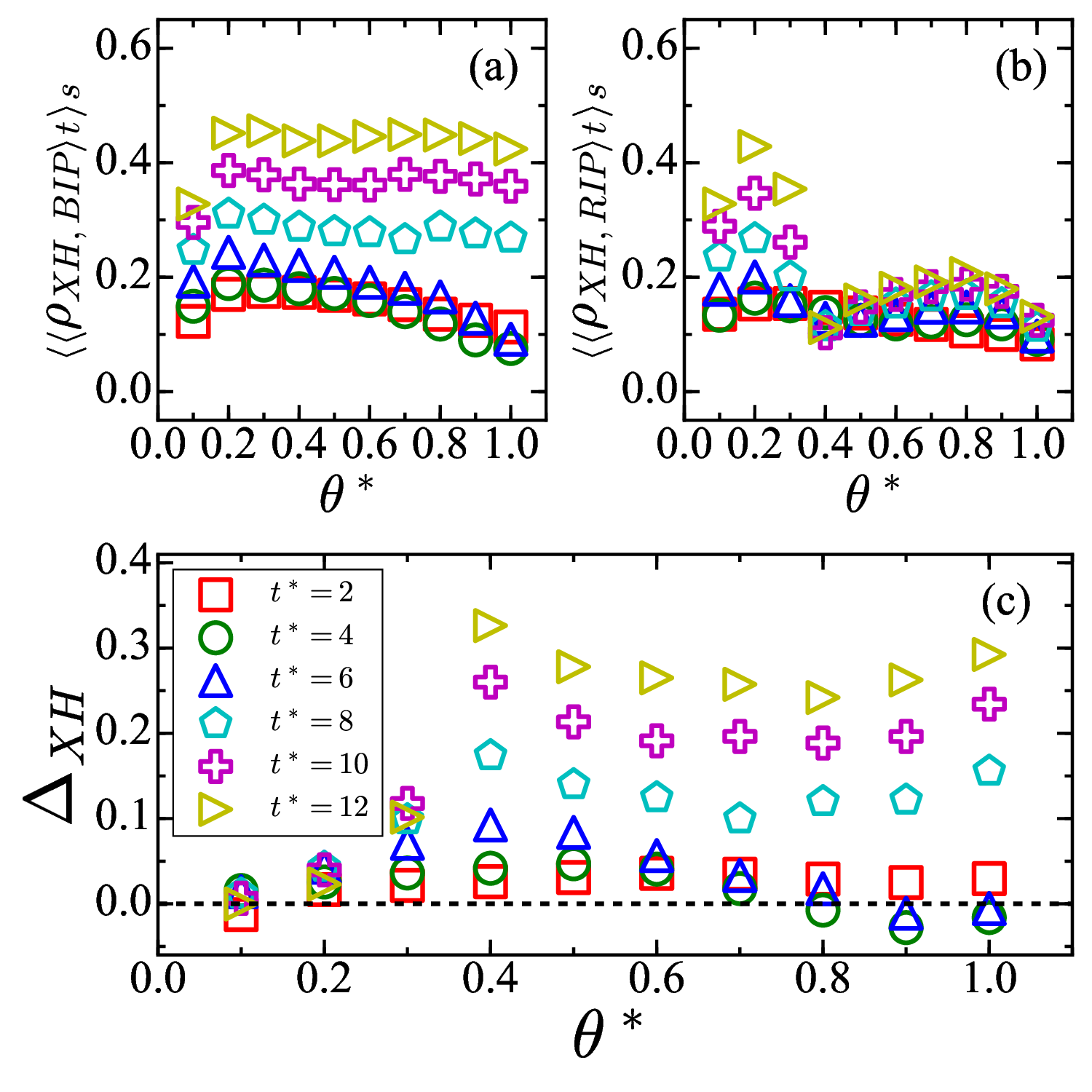}
    \caption{We plot the average fraction of the isolated nodes per unit time under the
    (a) BIP and (b) RIP for $t^*=2\sim 12$ and $\theta^*=0.1\sim 1.0$.
    Under the BIP, $\left<\left<\rho_{XH,BIP}\right>_t\right>_s$ hardly change as $\theta^*$ increases for a fixed $t^*$.
    However, when the RIP is applied, $\left<\left<\rho_{XH,RIP}\right>_t\right>_s$
    drastically decrease when $\theta^*>0.3$. and $t^*> 6$. This suggests that with high traceability,
    the RIP can effectively mitigate the epidemic spreading while minimizing the damage on
    the social and economic system.}
    \label{IsolPerTime}
\end{figure}
In epidemic control, reducing the number of isolated individuals at each time step becomes another crucial factor to minimize social and economic recession.
Here, we define the fractions of the isolated nodes per unit time
(with $\sigma_2=X$ or $H$) for protocol $Y$ as,
\begin{equation}
    \left< \rho_{XH,Y} \right>_{t} = 
    \frac{\int_{0}^{T_{final}} (\rho_{X}(t)+\rho_{H}(t)) dt}{T_{final}},
    \label{eq2}
\end{equation}
where $T_{final}$ represents the time at which $\rho_I(t)$ becomes zero.
In \FIGS{IsolPerTime} (a) and (b), we plot
$\left<\left<\rho_{XH,BIP}\right>_{t}\right>_{s}$
and $\left<\left<\rho_{XH,RIP}\right>_{t}\right>_{s}$, where
$\left<...\right>_{s}$ denotes the sample average over independent samples.
The sample averages are obtained from 500 independent trajectories.
The data in \FIG{IsolPerTime} (a) shows that 
$\left<\left<\rho_{XH,BIP}\right>_{t}\right>_{s}$ hardly changes as 
$\theta^{*}$ increases except for $\theta^{*} = 0.1$. Moreover, longer
isolation period leads to a larger values of 
$\left<\left<\rho_{XH,BIP}\right>_{t}\right>_{s}$ in general.
However, when the RIP is adopted, there is a significant drop in
$\left<\left<\rho_{XH,RIP}\right>_{t}\right>_{s}$ when $\theta^{*}>0.3$
and $t^{*}>6$.

For a quantitative analysis, we define the difference in the fractions of the isolated nodes per unit time between two protocols as 
\bea
\Delta_{XH}\equiv \left<\left<\rho_{XH,BIP}\right>_{t}\right>_{s}-\left<\left<\rho_{XH,RIP}\right>_{t}\right>_{s}.
\eea
By definition, if $\Delta_{XH}>0$ then more nodes are isolated under the BIP than the RIP.
As shown in \FIG{IsolPerTime} (c), for all values of $\theta^*$, we find 
that $\Delta_{XH}\approx 0$ for $t^*\le 6$. However, we find that
$\Delta_{XH}>0$ when $t^*>6$ and $\Delta_{XH}$ increases as $t^*$
increases. Thus, $t^*$ affects only $\rho_X$ and $\rho_H$ per unit time for both models.
Note that, even though $\Delta_{XH}\le 0$ for $t^*\le 6$,
$\Delta_R$ increases with $\theta^*$ and $\Delta_R\ge 0$
as shown in \FIG{FinalSize} (c) (see also \FIG{rhos}).
Therefore, the RIP more effectively controls the disease spreading 
through the preemptive isolation of suspicious contacts with fewer isolated nodes per unit time than the BIP.
We also display the measured $\Delta_R$ and $\Delta_{XH}$ when $P_F(k)\sim k^{-\gamma}$
in Appendix \ref{epidemicsize_sfn}, which are almost identical with those in \FIGS{FinalSize} 
and \ref{IsolPerTime}.
\section{Discussion}

In summary, we model the NPI adopted by many health authorities over the world, and
introduce a model for reinforced NPI.
In these models the state of each individual is characterized by three disease states
with additional isolation states.
Two different types of transmission routes observed in real world are implemented by
the multiplex networks.
By using numerical simulations, we compare the efficacy of the two models, BIP and RIP models,
and find that the RIP controls the spreading
of disease more efficiently by reducing both the epidemic size and the average number of isolated
individuals per unit time, despite its simplicity.
Especially, when the traceability is maximal, the final fraction of the susceptible nodes under the RIP
increases by almost 100\% (almost doubled) compared to that under the BIP.
This indicates that the RIP significantly and efficiently
protect the susceptible nodes through the preemptive isolation
of the possible contacts.
Furthermore, since we do not assume any characteristic property of a specific disease, we expect that
the suggested models can be used as a general framework for modeling disease control for any
real disease outbreak.

\begin{acknowledgments}
This research was supported by Basic Science Research Program
through the National Research Foundation of Korea (NRF) funded by
the Ministry of Education (Republic of Korea) 
(grant number: NRF-2019R1F1A1058549). 
\end{acknowledgments}

\appendix
\section{Epidemic Threshold on DLMN}
\label{epidemicthreshold}
\subsection{Estimating $\beta_c$}

To estimate $\beta_c$ we first define the transmissibility $\mathcal{T}$.
When a node $j$ is infected at time $t=t_0$, a susceptible neighbor $i$
can be infected by $j$ at time $t=t_0+n$ with the probability $\beta (1-\beta)^{n-1}$ where $n = 1, 2, ...,t_R$. By adding up the
probabilities for all possible values of $n$, we can get $\mathcal{T}$ as \cite{Lagorio11}
\bea
\mathcal{T} = \sum_{n=1}^{t_R}\beta(1-\beta)^{n-1}=1-(1-\beta)^{t_R}.
\label{transmissibility}
\eea
The transition between the disease-free phase and the epidemic phase is determined by the
average number of the secondary infections per infected node. The average number
of the secondary neighbor is
\bea
\sum_{k}k\frac{kP(k)}{\left< k \right>} - 1 = \frac{\left< k^2 \right>-\left< k \right>}{\left< k \right>}=\kappa-1,
\label{secondnn}
\eea
where $\kappa\equiv \left< k^2\right>/\left<k\right>$ is known as the {\it branching factor}.
Thus, when $\mathcal{T}(\kappa-1)\geq 1$, the epidemic spreads out over the network (epidemic phase). 
However, when $\mathcal{T}(\kappa-1)< 1$, the disease dies out in a short time scale (disease-free phase) \cite{Lagorio11}.
Therefore, the critical transmissibility $\mathcal{T}_c$ is given by 
\bea
 \mathcal{T}_c =\frac{1}{\kappa -1} =\frac{\left< k \right>}{\left< k^2 \right>-\left< k \right>}.
 \label{Tc}
\eea
From \EQS{transmissibility}{Tc}, $\beta_c$ is determined by the relation 
\bea
1-(1-\beta_c)^{t_R} = \frac{\left< k \right>}{\left< k^2 \right>-\left< k \right>}.
\label{beta_c}
\eea

\subsection{$\left<k\right>$ and $\left<k^2\right>$ of DLMN}
Let $G_F(x)$ and $G_W(x)$ be the generating functions of $P_F(k)$ and $P_W(k)$, respectively, which are defined by
\bea
G_F(x)=\sum_{k=0}^{\infty} P_F(k)x^k
\label{gf}
\eea
and
\bea
G_W(x)=\sum_{k=0}^{\infty} P_W(k)x^{k}.
\label{gw}
\eea
Then the combined degree distribution of both layers becomes
\bea
P(k^\prime)=\sum_{k_1=1}^{\infty}\sum_{k_2=1}^{\infty} \delta(k^\prime,k_1+k_2)P_F(k_1)P_W(k_2).
\eea
Here $\delta(i,j)$ is Kronecker's delta.
The generating function of $P(k^\prime)$, $G(x)$, is simply written as
\bea
G(x)=G_F(x)G_W(x).
\label{g}
\eea

On DLMN due to the time-dependent feature of degree on the $W$-layer,
$P_W(k)$ is well approximated by the Poisson distribution,
$P_W(k)=\left<k\right>^{-k} \exp(-\left<k\right>)/k!$.
When $P_F(k)$ is given by $P_F(k)=\left<k\right>^{-k} \exp(-\left<k\right>)/k!$,
from Eq. \ref{gf}, \ref{gw} and \ref{g}, we obtain
\bea
G(x)=\exp\left\{ 2\left<k\right> (x-1)\right\},
\eea
and $\left<k^\prime\right>=2\left<k\right>$ and
$\left<{k^\prime}^2\right>=2\left<k\right>+(2\left<k\right>)^2$.
Thus, we numerically estimate the threshold from \EQ{beta_c} as $\beta_c\approx 0.0107$.

On the other hand, when $P_F(k)=k^{-\gamma}/\zeta(\gamma)$ and $P_W(k)=\left<k\right>^{-k} \exp(-\left<k\right>)/k!$,
we obtain $\left<k^\prime\right>=(\left<k\right> \mathrm{Li}_{\gamma}(1)+\mathrm{Li}_{\gamma-1}(1))/\zeta(\gamma)$
and $\left<{k^\prime}^2\right>=\left[\left<k\right>(\left<k\right>+1)\mathrm{Li}_{\gamma}(1)+2\left<k\right>\mathrm{Li}_{\gamma-1}(1)+\mathrm{Li}_{\gamma-2}(1)\right]/\zeta(\gamma)$.
Here $\zeta(\gamma)$ is the Riemman zeta function and $\mathrm{Li}_{\gamma}(x)$ is
the polylogarithm of $x$. Thus, $\beta_c=0$ for $2<\gamma<3$. 

Therefore, $\beta=0.2$ guarantees that the the system is in the epidemic phase,
regardless of the structure of $F$-layer with given parameters.

\section{Oscillatory Behavior}
\label{Oscillatory}
\begin{figure}[h]
    \centering
    \includegraphics[width=0.4\textwidth]{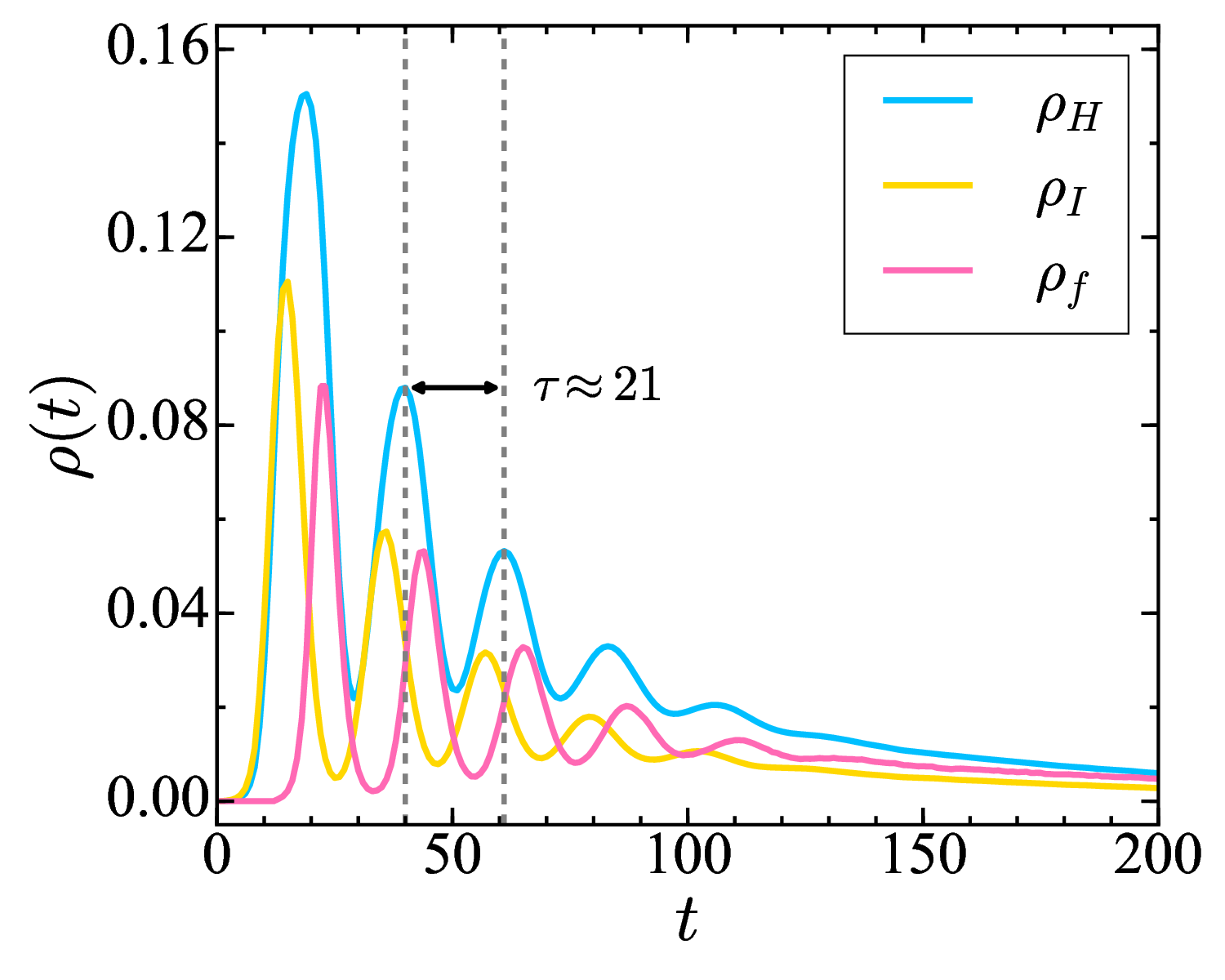}
    \caption{Plot of the fractions $\rho_{H}$, $\rho_{I}$ and $\rho_{f}$
    under the RIP. 
    The epidemic parameters are $\beta=0.2$ and $t_{R}=6$,
    and the intervention parameters are $\theta^*=0.9$ and $t^*=12$.
    }
    \label{oscillation}
\end{figure}

Under the RIP with large values of $\theta^{*}$ and $t^{*}$, oscillatory
behaviors are observed in the fraction of nodes in each sate when $P_F(k)$ follows the Poisson distribution.
In \FIG{oscillation}, we plot $\rho_H$, $\rho_I$ and $\rho_f$ 
which clearly shows such oscillatory behaviors. 
Here, $\rho_{f}(t)$ is  defined as the fraction of the susceptible nodes which are just released from self-isolation at time $t$.
We used the intervention parameters as $\theta^{*} = 0.9$ and $t^{*} = 12$. The period of oscillation $\tau$ for each fraction with the given 
parameters  is estimated as $\tau \approx 21$. The peak position of 
each curve indicates that the increase of the infected
causes an increase of the hospitalized individuals. 
Due to the hospitalization and self-isolation,
the number of the infected rapidly decreases. 
However, after $t^*$ the isolated
nodes are set to be free which increases the number of unisolated susceptible nodes.
Thus it increases the number of infected individuals again.
This pattern is repeated with decreasing amplitude due to the depletion of the
susceptible nodes until there is no more infected node left.
This oscillatory behavior is observed only for the case of large $\theta^*$ and $t^*$ in the RIP.

\section{Results when $P_F(k)\sim k^{-\gamma}$}
\label{sfn}
In this section we summarize the obtained results when $P_F(k)$ follows
the power-law, $P_F(k)\sim k^{-\gamma}$, with degree exponent $\gamma=2.7$.
For a direct comparison with the results in the main text, we set $\left<k\right>=8$
and  $N=100,000$. We use the static model \cite{Goh01} to construct the $F$-layer with a power-law degree distribution.
Except the underlying topology of $F$-layer, other parameters
are the same with those in the main text.

\subsection{Evolution of $\left\{\rho_m(t)\right\}$'s}
\label{rho_sfn}
\begin{figure*}[ht]
    \centering
    \includegraphics[width=0.9\textwidth]{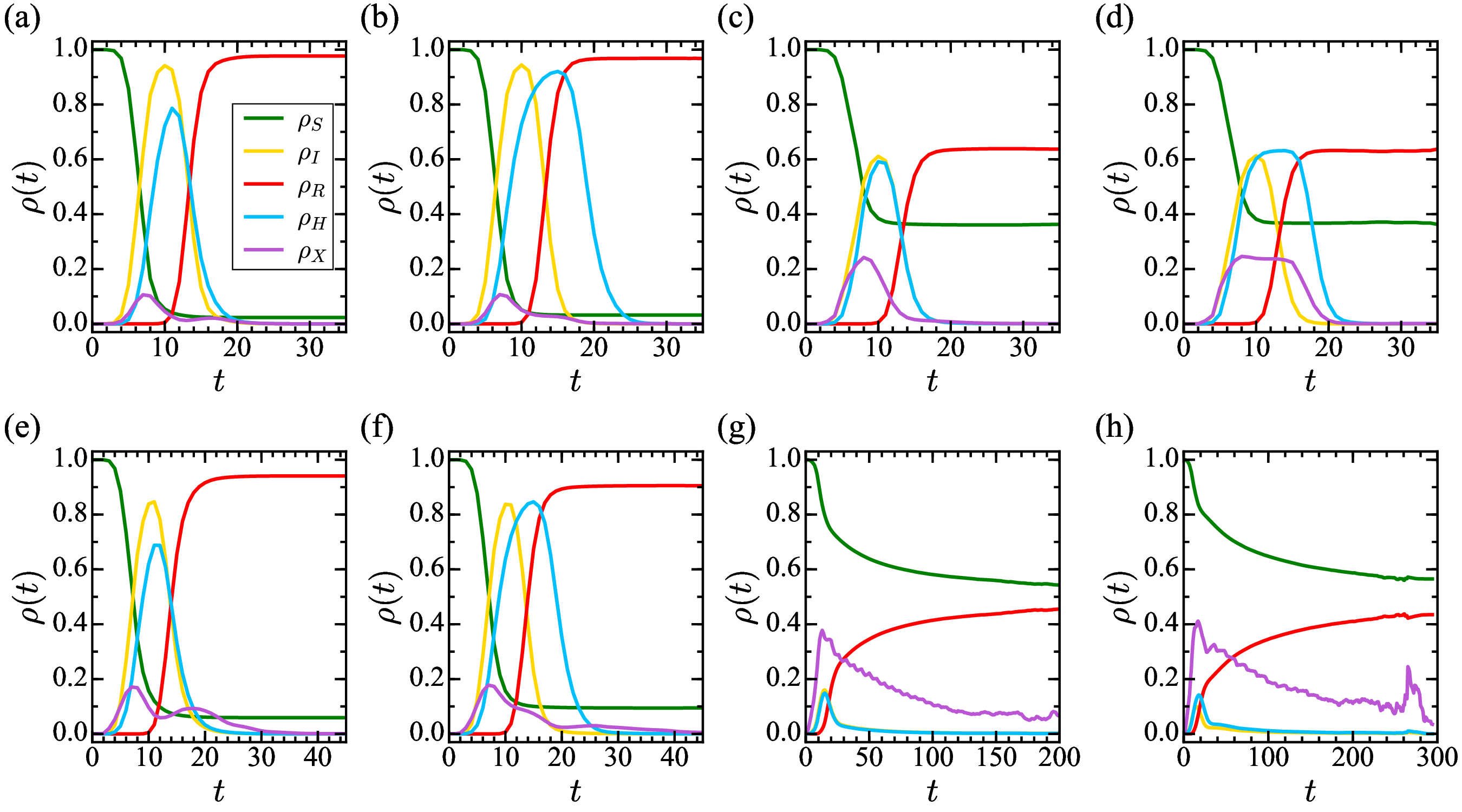}
    \caption{(a)-(d) are \{$\rho_m(t)$\}'s ($m\in \left\{S,I,R,H,X\right\}$) under the BIP
    when $P_{F}(k)\sim k^{-\gamma}$
    with (a) $\theta^*=0.3$, 
    $t^*=4$, (b) $\theta^*=0.3$, $t^*=10$, (c) $\theta^*=0.9$, $t^*=4$,
    (d) $\theta^*=0.9$, $t^*=10$. 
    (e)-(h) shows \{$\rho_m(t)$\}'s under the RIP with 
    (e) $\theta^*=0.3$, $t^*=4$, (f) $\theta^*=0.3$, $t^*=10$,
    (g) $\theta^*=0.9$, $t^*=4$, (h) $\theta^*=0.9$, $t^*=10$.
    $\beta=0.2$ and $t_{R}=6$ are used in common.
    Each data is obtained from 500 independent simulations by
    averaging the surviving samples at time $t$.}
    \label{rhos-sf}
\end{figure*}

The data in \FIG{rhos-sf} shows $\left\{\rho_m(t)\right\}$'s when  $P_F(k)\sim k^{-\gamma}$.
As shown in \FIG{rhos-sf}, the qualitative behavior of $\left\{\rho_m(t)\right\}$'s are
almost the same with those in  \FIG{rhos}. The only difference is the decrease of
$\rho_R$ when $\theta^*$ is large (see \FIGS{rhos-sf}(g) and (h)). 
Since the average number of the secondary neighbors on $F$-layer becomes large if
$P_F(k)$ follows the power-law, more nodes are in the $X$ state compared with 
the Poisson distribution case. This effect becomes larger as $\theta^*$ increases.
Thus, more nodes are isolated and protected from the infection. As a result, the
$\rho_S$ increase when $P_F(k)\sim k^{-\gamma}$ as shown in \FIGS{rhos-sf}(g) and (h).

\subsection{Final epidemic size and the number of isolated nodes per unit time}
\label{epidemicsize_sfn}
\begin{figure}[h]
    \centering
    \includegraphics[width=0.4\textwidth]{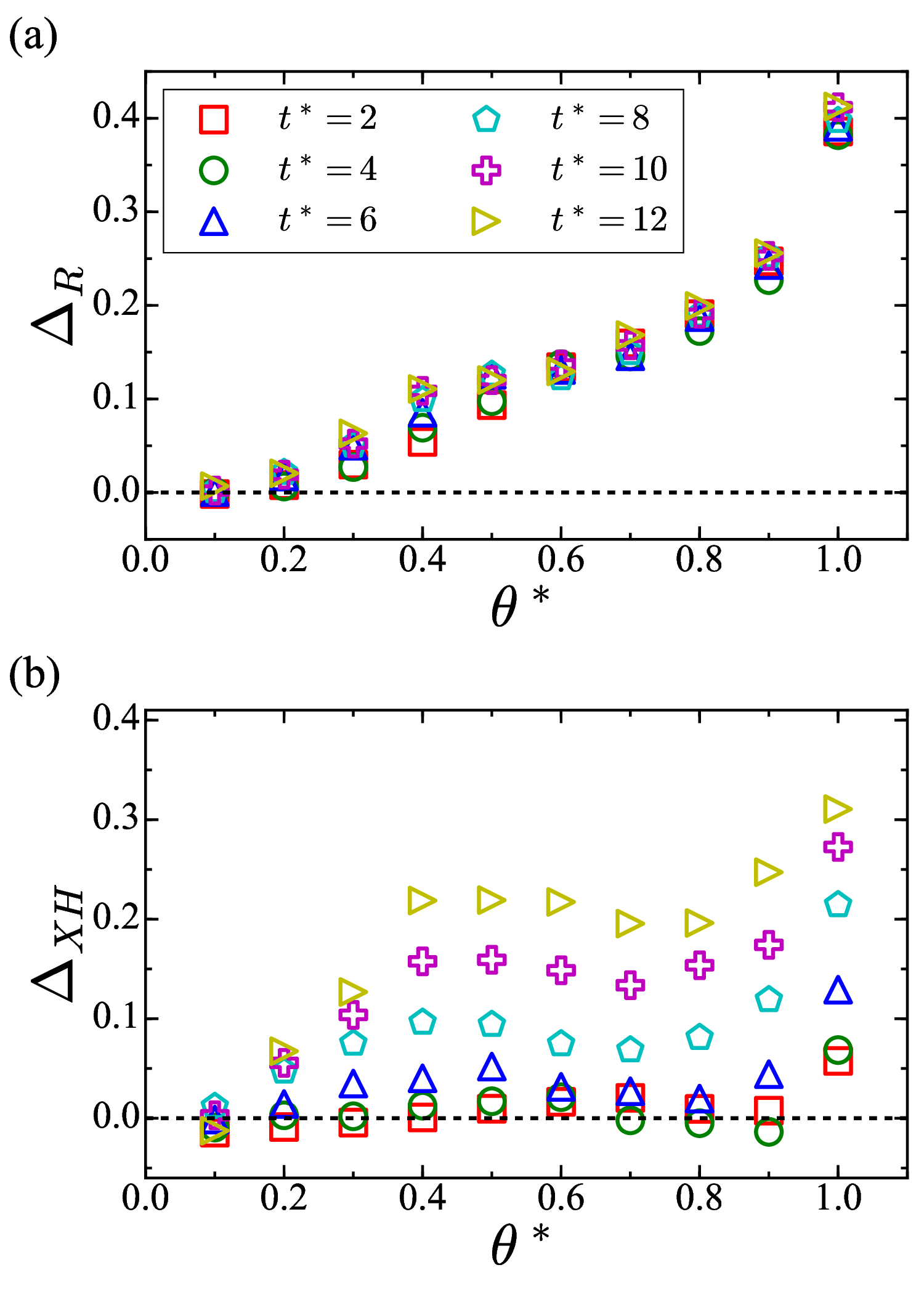}
    \caption{We plot
    (a) the difference of the final epidemic sizes between the BIP and RIP, $\Delta_{R}$ and
    (b) the difference of the numbers of isolated nodes per unit time between the BIP and RIP, $\Delta_{XH}$ when $P_{F}(k)\sim k^{-\gamma}$.
    The black horizontal dashed line denotes
    $\Delta_{R}=0$.}
    \label{Delta-sf}
\end{figure}

In \FIG{Delta-sf} we also display the measured difference of the
final epidemic sizes and the numbers of isolated nodes between the BIP and RIP for various values of $\theta^*$.
As shown in the data, we find that the underlying topology of $F$-layer hardly affect
the behavior of final epidemic size and number of isolated nodes. The obtained
data are qualitatively the same with those in Figs.\ref{FinalSize} (c) and \ref{IsolPerTime} (c). 

%

\end{document}